\documentclass[reqno,11pt]{amsart} 

\usepackage{a4}
\usepackage{epsf}

\usepackage{amssymb}
\usepackage{cite}
\newcommand{\beq}{\begin{equation}}
\newcommand{\ee}{\end{equation}}
\newcommand{\bea}{\begin{eqnarray}}
\newcommand{\eea}{\end{eqnarray}}

\usepackage{hyperref}
\def\stackreb#1#2{\ \mathrel{\mathop{#1}\limits_{#2}}}

\newcommand{\CC}{\mathbb C}

\newcommand{\Z}{\mathbb Z}

\newcommand{\ve}{\varepsilon}

\begin{document}

\title[General modular quantum dilogarithm and beta integrals]
{General modular quantum dilogarithm  \\[0.2em]
and beta integrals}

\author{Gor A. Sarkissian and Vyacheslav P. Spiridonov}
 \address{
Laboratory of Theoretical Physics, JINR, Dubna, 141980,  Russia;
Department of Physics, Yerevan State University, Alex Manoogian 1, 0025, Yerevan, Armenia;
National Research University Higher School of Economics, Moscow, Russia}

\begin{abstract}
We consider a univariate beta integral composed from general modular quantum dilogarithm functions
and prove its exact evaluation formula. It represents the partition function of a particular
$3d$ supersymmetric field theory on the general squashed lens space.
Its possible applications to $2d$ conformal field theory are briefly discussed as well.
\end{abstract}

\maketitle

\tableofcontents

\begin{flushright}
To A.~A. Slavnov on the occasion of his 80th birthday
\end{flushright}

\section{Introduction}

For many generations of students in Russia, including the authors, the basic sources of knowledge
on quantum field theory were the textbook by Bogoliubov and Shirkov \cite{BSh} and the monograph
of Slavnov and Faddeev \cite{SF}, complementing the fundamental treatise \cite{BSh}
by a consideration of the gauge fields theory.
However, nowadays they do not form a complete background for entering a modern research.
One of the most important missing ingredients in these books is supersymmetry,
which became vital for understanding perturbative and non-perturbative
theoretical mechanisms behind some physical phenomena. In particular, the last three decades
saw the rise of the highly powerful localization technique allowing for exact computation
of various partition functions (including the superconformal indices) of
supersymmetric field theories on curved space-times \cite{local}.

These partition functions are expressed in terms of the complicated special functions
of hypergeometric type. Trying to understand their properties, it is natural to ask --
whether they are new or not? Clearly they are beyond the existing textbooks
like \cite{aar} and handbooks like \cite{dlmf} and require searching in the
modern mathematical literature. Raising such a question, Dolan and Osborn
have found in 2008 \cite{DO} a direct connection of superconformal indices of $4d$ $\mathcal{N}=1$
theories to the elliptic hypergeometric integrals constructed eight years earlier \cite{spi:umn}.
The discovery of elliptic hypergeometric functions at the turn of millenium
became a big surprise to mathematicians, because the theory of special functions
of hypergeometric type has been developed since the times of Euler only in two instances --
the plain hypergeometric functions, like the Euler-Gauss ${}_2F_1$-function, and their
$q$-analogues \cite{aar}, and there were no indications on the existence of the third
level for such functions.
Moreover, these new functions unified two previously separately considered families of classical
special functions (elliptic and hypergeometric ones) and generalized all previously found
hypergeometric objects \cite{spi:essays}.

The connection with quantum field theory appeared to be very fruitful, since it resulted
in new understanding of the structure of these functions and brought many new both
mathematical and physical results (see, e.g. \cite{SV1} or \cite{local}).
The present paper represents another step in the development of such relations.
Namely, we describe an extension of $q$-hypergeometric functions constructed from
the modular quantum dilogarithm. Modular analogue of the quantum dilogarithm function was
suggested by Faddeev in connection to the lattice Virasoro algebra \cite{Fad94,Fad95}, and it has found many
applications, in particular, in the hyperbolic Ruijsenaars model \cite{Ruij} and relativistic
Toda chain \cite{KLS}, Yang-Baxter equation \cite{BMS,BT06,CD14,kashaev,VF},
topological invariants \cite{kashaev2,Dimofte:2011ju},
usual $2d$ conformal field theory \cite{FZZ} and its discretizations \cite{BMS2}.
In the relatively recent time it was found to play a key role in the computations
of the supersymmetric partition functions of $3d$ models \cite{KWY} on the squashed
three-sphere $S^3_\tau$ (see \cite{HHL}).

In the theory of special functions, a slight modification of Faddeev's function was called
the hyperbolic gamma function \cite{Ruij}, and the integrals composed out of them were
called the hyperbolic hypergeometric integrals. The top univariate hyperbolic beta integral
was described in \cite{stok} and it represented a special degeneration of the
elliptic beta integral \cite{spi:umn}, or, more generally, of the elliptic analogue
 of the Euler-Gauss hypergeometric function \cite{spi:essays}, as described in \cite{BRS}.
This integral has been directly connected to the functional star-triangle relation \cite{spi:conm}
and the Yang-Baxter equation \cite{CS}, supersymmetric partition function of $3d$ theories \cite{DSV},
 and a topological field theory \cite{kashaev3}.

Faddeev's function used the simplest $\tau\to -1/\tau$  transformation from the modular group
$SL(2,\mathbb{Z})$ applied to the infinite product $(z;q)_\infty=\prod_{j=0}^\infty(1-zq^n)$,
$q=e^{2\pi i\tau}$. Its generalization to a function based on an arbitrary modular transformation
from $SL(2,\mathbb{Z})$ was suggested by Dimofte in \cite{dimofte}. The key motivation was an
interest in the partition function of the Chern-Simons theory on the general squashed lens space
$L(c,a)_\tau$. The $a=-1$ case of this function was considered earlier in \cite{Imamura:2012rq}
and around the same time in \cite{AP,AK,NP}. $q$-Hypergeometric functions associated with $L(c,-1)_\tau$
were investigated in \cite{GK,spi:rare}. A manifestation of the corresponding hyperbolic integrals
in $2d$ conformal field theory was discussed in \cite{SS}.

Dimofte proved a pentagon relation for the general modular quantum dilogarithm,
which can be considered as a special identity for the corresponding generalized hyperbolic
hypergeometric integral. Its special case associated with the manifold $S^3_\tau/\mathbb{Z}_k$
was independently established in \cite{AK}.
Following the terminology suggested in \cite{spi:rare}, we will be calling also the function
introduced in \cite{dimofte} as the rarefied hyperbolic gamma function. Taking it as a building
block, we construct the univariate hyperbolic beta integral associated with $L(c,a)_\tau$
and prove its exact evaluation formula. This integral represents the partition function
of a particular $3d$ supersymmetric field theory and defines a new class of solutions of
the star-triangle relation. In particular, this formula and its degenerations extend the
considerations of \cite{GK,SS}. Also, the derived formula should
be applicable to the general lens space extension of $2d$ conformal field theory
considered in \cite{Bonelli}, similarly to the parafermionic case \cite{SS}.
The main result of the present work was announced in a short note \cite{SS2}.

\section{The modular quantum dilogarithm}

Let us take $SL(2,\Z)$ group of modular transformations described by the matrices
\beq
M=\left(
\begin{array}{cc}
a & b  \\
c & d
\end{array} \right), \quad ad-bc=1, \quad a,b,c,d\in\Z.
\ee
Its projective realization allowing simultaneous change of the signs of all integer parameters,
\begin{equation}
\tau \to \frac{a\tau +b}{c\tau+d},
\label{PSL}\end{equation}
plays an important role in the theory of automorphic forms. In the following we use only
transformation \eqref{PSL} with the fixed sign of $c$, $c\geq 0$,

In order to match with the notations of \cite{dimofte}, we also set
\beq
a= -p, \quad b= -s, \quad c=k,\quad d=-r, \quad k>0,\quad pr+ks=1.
\ee
Since $k$ will play a special role, we write also $pr=1-ks=1\, \textrm{mod}\; k$,
so that  $(p|s)=(p|k)=(r|k)=(r|s)=1$ (here $(p|k)$ denotes the greatest common devisor of $p$ and $k$).

The three-dimensional sphere $S^3$ can be described by two complex variables $z_1$ and $z_2$
satisfying the equation $|z_1|^2+|z_2|^2=1$. Its squashed form $S^3_\tau$
is defined by the relation $|\kappa z_1|^2+|\kappa^{-1} z_2|^2=1$, where $\kappa=\kappa(\tau)$
is some deformation parameter. The squashed lens space $L(c,a)_\tau$ is defined by the same relation
 after the following identification of points \cite{lens}
\beq\label{lens_space}
(z_1,z_2)\sim (e^{\frac{2\pi i}{c}}z_1,e^{\frac{2\pi i a}{c}}z_2), \quad a\in \Z_{>0},\quad (a|c)=1.
\ee
In this context, because of the mod $c$ restrictions, one has an additional reduction of the modular group
parameters -- there remain only two integer variables $a$ and $c$.

Following Dimofte \cite{dimofte}, we define the general unnormalized modular quantum dilogarithm,
or the rarefied hyperbolic gamma function, as
\beq\label{hyplens}
\gamma_M(\mu,m)=\gamma_M(\mu,m;\omega_1,\omega_2):=\frac{(\tilde q e^{2\pi i \tilde u(\mu, m)};\tilde q)_\infty}
{(e^{2\pi i u(\mu,m)};q)_\infty}, \quad |q|<1,
\ee
where $(a;q)_\infty=\prod_{j=0}^\infty (1-aq^j)$,
\beq
u(\mu,m):=\frac{\mu+m\omega_2}{k\omega_2}, \quad q:=e^{2\pi i \tau},\quad \tau:=\frac{\omega_1+r\omega_2}{k\omega_2},
\ee
and
\beq
\tilde q:=e^{2\pi i\tilde \tau}, 
\quad \tilde \tau:=\frac{a\tau+b}{c\tau+d}
=-\frac{p\omega_1+\omega_2}{k\omega_1},
\ee
together with
\beq
 e^{2\pi i\tilde u(\mu,m)}=e^{2\pi i\frac{\mu-pm\omega_1}{k\omega_1}}
=\tilde q^m e^{\frac{2\pi i u(\mu,m)}{c\tau+d}}, \quad c\tau+d=\frac{\omega_1}{\omega_2}.
\ee
Because of the evident periodicity $\gamma_M(\mu,m+k)=\gamma_M(\mu,m)$ we shall
assume that $m\in \Z_k,\, \Z_k=\{0,1,\ldots, k-1\}$.
As argued in \cite{dimofte}, $\gamma_M(\mu,m)$ is a general lens space  analogue of Faddeev's
quantum modular dilogarithm \cite{Fad94,Fad95}. We shall call it also the rarefied hyperbolic gamma function.
This function, after appropriate normalization, gives rise to a compact form of the
general rarefied hyperbolic beta integral evaluation formula, which is the main desired goal of this work.

Let us denote $2\pi i \omega_1/\omega_2=-\delta$ and take the limit $\delta\to 0^+$.
Then $q\to \epsilon,\, \epsilon=e^{2\pi ir/k},\, \epsilon^k=1$ and $\tilde q\to 0$.
As a result we have the asymptotics
\beq
\gamma_M(\mu,m)=\exp \sum_{n=1}^\infty \Big( \frac{e^{2\pi i n u }}{n(1-q^n)}
-\frac{\tilde q^n e^{2\pi i n \tilde u }}{n(1-\tilde q^n)} \Big)
\stackreb{\propto}{\delta\to 0} \exp\Big(\frac{1}{k\delta} \textrm{Li}_2(e^{2\pi i\frac{\mu}{\omega_2}})\Big),
\ee
where $\textrm{Li}_2(x)=\sum_{n=1}^\infty\frac{ x^n}{n^2}$ is the dilogarithm function.
This relation justifies the name {\em modular quantum dilogarithm} for $\gamma_M(\mu,m)$,
as suggested by Faddeev for the $k=p=r=1$ case in \cite{Fad95}.

The infinite product standing in the denominator of (\ref{hyplens}) has zeros at the points
$$
\mu=- j\omega_1- (k n+m+jr)\omega_2,\quad n\in\Z,\quad j\in\Z_{\geq0}.
$$
The numerator function has zeros at the points
$$
\mu= (k n'+p(m+ j'+1))\omega_1+(j'+1)\omega_2,\quad n'\in\Z,\quad j'\in\Z_{\geq0}.
$$
There are coinciding points in these two sets, namely, the zeros with the coordinates
$$
k n+m+j'+1+rj=0, \quad k n'+p(m+ j'+1)+j=0
$$
cancel each other.
Multiplying the first equation by $p$ and subtracting the second one we find
$$
n'=pn-sj, \quad kn+rj+m=-1,-2,\ldots.
$$
Therefore true poles of function \eqref{hyplens} are located at the points
\beq
\mu=- j\omega_1- (k n+rj+m)\omega_2,\quad n\in\Z,\quad j\in\Z_{\geq0},
\quad kn+rj+m\in \Z_{\geq 0},
\ee
and true zeros lie at the points
\beq
\mu= (p(m+ j+1)+k n)\omega_1+(j+1)\omega_2,\quad n\in\Z,\quad j\in\Z_{\geq0},
\quad p(m+ j+1)+k n\in \Z_{>0}.
\ee

Relations
\bea\nonumber &&
e^{2\pi i u(\mu+\omega_1,m+r)}=q e^{2\pi i u(\mu,m)}, \quad
e^{2\pi i u(\mu+\omega_2,m-1)}=e^{2\pi i u(\mu,m)},
\\ &&
e^{2\pi i \tilde u(\mu+\omega_1,m+r)}=e^{2\pi i \tilde u(\mu,m)}, \quad
e^{2\pi i \tilde u(\mu+\omega_2,m-1)}=\tilde q^{-1} e^{2\pi i \tilde u(\mu,m)}
\nonumber\eea
lead to the equations
\bea &&
\gamma_M(\mu+\omega_1,m+r)=(1-e^{2\pi i \frac{\mu+m\omega_2}{k\omega_2}})\gamma_M(\mu,m),
\\ &&
\gamma_M(\mu+\omega_2,m-1)=(1-e^{2\pi i \frac{\mu-pm\omega_1}{k\omega_1}})\gamma_M(\mu,m).
\eea
Making back shifts of the variables, we obtain
$$
\gamma_M(\mu-\omega_1,m-r)=\frac{\gamma_M(\mu,m)}
{1-q^{-1}e^{2\pi i \frac{\mu+m\omega_2}{k\omega_2}}}, \quad
\gamma_M(\mu-\omega_2,m+1)=\frac{\gamma_M(\mu,m)}
{1-\tilde q e^{2\pi i \frac{\mu-pm\omega_1}{k\omega_1}}}.
$$

For $k=p=r=1, s=0$, i.e. $M=M_{st}: =\tiny\left(
\begin{array}{cc}
-1 & 0  \\
1 & -1
\end{array} \right)$, the dependence on $m$ disappears and one gets the standard Faddeev's
modular quantum dilogarithm (hyperbolic gamma function)
\beq\label{gamone}
\gamma(\mu)=\gamma(\mu;\omega_1,\omega_2):=\gamma_{M_{st}}(\mu,m)
=\frac{(\tilde q e^{2\pi i \frac{\mu}{\omega_1}};\tilde q)_\infty}
{(e^{2\pi i \frac{\mu}{\omega_2}};q)_\infty}, \quad
q= e^{2\pi i\frac{\omega_1}{\omega_2}}, \quad  \tilde q= e^{-2\pi i\frac{\omega_2}{\omega_1}}.
\ee
Using the identity $ (a;q)_{\infty}=\prod_{n=0}^{k-1}(aq^n;q^k)_{\infty}, $
one can write the function $\gamma_M(\mu,m)$ as the following product \cite{dimofte}
\beq\label{gampk}
\gamma_M(\mu,m)=\prod_{\gamma,\delta \in \Delta(k,p,m)} \gamma\left({1\over k}(\mu+\omega_1\delta+\omega_2\gamma);\omega_1,\omega_2\right),
\ee
where $\Delta(k,p,m)=\{\gamma,\delta\in\mathbb{Z}, 0\leq \gamma,\delta< k, p\gamma-\delta \equiv pm\, {\rm mod}\, k$\}.
Multiplying the constraint $p\gamma-\delta \equiv pm\; {\rm mod}\, k$ by $r$, we can write it also as $\gamma-r\delta \equiv m
\, {\rm mod}\, k$.

Using expressions (\ref{hyplens}) and (\ref{gamone}), for general
$M_1=\tiny\left(
\begin{array}{cc}
-p & -s  \\
k & -r
\end{array} \right)$,
$M_2=\tiny\left(
\begin{array}{cc}
r & k  \\
-s & p
\end{array} \right)$,
$M_3=\tiny\left(
\begin{array}{cc}
p & -k  \\
s & r
\end{array} \right)$
it is straightforward to derive the identities
\bea && \nonumber
\gamma_{M_1}(\mu,m;\omega_1,\omega_2)= \gamma_{M_2}(-s(\mu+m\omega_2),0;p\omega_1+\omega_2,r\omega_2+\omega_1)
\\  \label{dual1}  && \makebox[2em]{} \times
\gamma\left({\mu+(k+m)\omega_2\over k};{r\omega_2+\omega_1\over k},\omega_2\right)
\gamma\left({\mu+(k-pm)\omega_1\over k};\omega_1,{\omega_2+p\omega_1\over k}\right),
\\ \nonumber &&
\gamma_{M_1}(\mu,m;\omega_1,\omega_2)
\gamma_{M_3}(s(\mu+m\omega_2),0;r\omega_2+\omega_1,p\omega_1+\omega_2)
\\ && \makebox[2em]{}
=\gamma\left({\mu+m\omega_2\over k};{r\omega_2+\omega_1\over k},\omega_2\right)
\gamma\left({\mu+(-pm)\omega_1\over k};\omega_1,{\omega_2+p\omega_1\over k}\right).
\label{dual2} \eea
It follows from (\ref{dual1}) that for $s=-1$, i.e. for $rp=1+k$, the function $\gamma_M(\mu,m)$ can be written as a product
of three $\gamma(\mu)$ functions:
 \bea\label{dual11} &&
 \gamma_{\tiny\left(
\begin{array}{cc}
-p & 1  \\
k & -r
\end{array} \right)}(\mu,m;\omega_1,\omega_2)=\gamma\left({\mu+(k+m)\omega_2\over k};{r\omega_2+\omega_1\over k},\omega_2\right)
\\ \nonumber && \makebox[2em]{} \times
\gamma\left({\mu+(k-pm)\omega_1\over k};\omega_1,{\omega_2+p\omega_1\over k}\right)
\gamma(\mu+m\omega_2;p\omega_1+\omega_2,r\omega_2+\omega_1).
\eea
Similarly, for the case $s=1$, i.e. for $rp=1-k$, $\gamma_M(\mu,m)$ can be expressed via
 three $\gamma(\mu)$ functions  as well,
\bea\label{dual12}
\gamma_{\tiny\left(
\begin{array}{cc}
-p & -1  \\
k & -r
\end{array} \right)}(\mu,m;\omega_1,\omega_2)
= {\gamma\left({\mu+m\omega_2\over k};{r\omega_2+\omega_1\over k},\omega_2\right)
\gamma\left({\mu+(-pm)\omega_1\over k};\omega_1,{\omega_2+p\omega_1\over k}\right)\over
\gamma(\mu+m\omega_2;r\omega_2+\omega_1,p\omega_1+\omega_2)}.
\eea
It is instructive to compare the derived formulae with the one corresponding to parafermionic functions
emerging for $p=r=1$ and $s=0$ \cite{Imamura:2012rq,NP,GK,SS}:
\bea\label{dual3} &&  \makebox[-3em]{}
\gamma_{\tiny\left(
\begin{array}{cc}
-1 & 0  \\
k & -1
\end{array} \right)}(\mu,m;\omega_1,\omega_2)=\gamma\left({\mu+m\omega_2\over k};{\omega_2+\omega_1\over k},\omega_2\right)
\gamma\left({\mu+(k-m)\omega_1\over k};\omega_1,{\omega_2+\omega_1\over k}\right).
\eea

Let us take definition of the Dedekind $\eta$-function and Jacobi $\theta_1$-function
\bea \label{eta} &&
\eta(\tau)=e^{\frac{\pi i\tau}{12}}(e^{2\pi i\tau};e^{2\pi i\tau})_\infty,
\\  &&
 \theta_1(u|\tau)=-\theta_{11}(u)=-\sum_{\ell\in \mathbb Z+1/2}e^{\pi {i} \tau \ell^2}
e^{2\pi {i} \ell (u+1/2)}
={i}q^{1/8}e^{-\pi {i} u}(q;q)_\infty\theta(e^{2\pi {i} u};q),
\label{theta1}\end{eqnarray}
with $q=e^{2\pi {i} \tau}$ and $\theta(z;q)=(z;q)_\infty (qz^{-1};q)_\infty.$
Modular transformation laws for $c>0$ have the form  \cite{rad}
\beq
\eta\left(\frac{a\tau+b}{c\tau+d}\right)=\ve(a,b,c,d)\sqrt{-i(c\tau+d)}\eta(\tau),
\ee
and
\beq
\theta_1\left(\frac{u}{c\tau+d}\Big|\frac{a\tau+b}{c\tau+d}\right)=-i \ve(a,b,c,d)^3\sqrt{-i(c\tau+d)} e^{\frac{\pi icu^2}{c\tau+d}}\theta_1(u|\tau),
\ee
where the character (24-th root of unity)
\beq
\ve=\ve(a,b,c,d):=
\left\{
\begin{array}{cl}
\left(\frac{d}{c}\right) e^{\frac{\pi i(1-c)}{4}} e^{\frac{\pi i}{12}[bd(1-c^2)+c(a+d)]}&
\textrm{for odd}\, c,
\\
\left(\frac{c}{d}\right) e^{\frac{\pi id}{4}} e^{\frac{\pi i}{12}[ac(1-d^2)+d(b-c)]} &
\textrm{for odd}\, d.
\end{array}
\right.
\ee
Here $\left(\frac{d}{c}\right)$ is the Legendre-Jacobi symbol. For odd $c>0$
$$
\left(\frac{d}{c}\right)=(-1)^{g_c(d)},\quad g_c(d)= \sum_{\nu=1}^{(c-1)/2}\left[\frac{2d\nu}{c}\right],
$$
where $[x]$ is the integer part of $x\in\mathbb{R}$.

Using the triple product Jacobi identity, we deduce the modular transformation rule for the $\theta(z;q)$-function:
\beq
\theta(e^{-\frac{2\pi i u}{c\tau+d}};e^{2\pi i\frac{a\tau +b}{c\tau+d}})
=i\ve^2
\, e^{\frac{\pi i}{6}(\tau-\frac{a\tau+b}{c\tau+d})}\,
e^{-\pi iu(1+\frac{1}{c\tau+d})}\,
e^{\frac{\pi icu^2}{c\tau+d}}\, \theta(e^{2\pi i u};e^{2\pi i\tau}).
\ee
Now we can deduce the reflection formula for our rarefied hyperbolic gamma function
\bea \nonumber &&
\gamma_M(\omega_1+\omega_2-\mu,r-1-m)\gamma_M(\mu,m)
=\frac{(\tilde q^{-m}e^{-\frac{2\pi iu}{c\tau+d}};\tilde q)_\infty
(\tilde q^{m+1}e^{\frac{2\pi iu}{c\tau+d}};\tilde q)_\infty}
{(qe^{-2\pi iu};q)_\infty(e^{2\pi iu}; q)_\infty}
\\ \nonumber  && \makebox[2em]{}
=(-1)^m\tilde q^{-\frac{m(m+1)}{2}}e^{-\frac{2\pi ium}{c\tau+d}}\, \frac{\theta(e^{-\frac{2\pi i u}{c\tau+d}};e^{2\pi i\frac{a\tau +b}{c\tau+d}})}
{\theta(e^{2\pi i u};e^{2\pi i\tau})}
\\ \nonumber &&  \makebox[2em]{}
=i\ve^2 e^{\frac{\pi i(r+p-3)}{6k}}e^{\pi i (1-s)m}e^{\pi i \frac{p}{k}m(m-r+1)}
e^{\frac{\pi i}{k}B_{2,2}(\mu;\omega_1,\omega_2)},
\eea
where
\beq
B_{2,2}(\mu;\omega_1,\omega_2)=\frac{1}{\omega_1\omega_2}
\left((\mu-\frac{\omega_1+\omega_2}{2})^2-\frac{\omega_1^2+\omega_2^2}{12}\right)
\ee
is the second order multiple Bernoulli polynomial.
Note that the right-hand side expression is invariant with respect to the shift $m\to m+k$.

For $k=p=r=1$ one has $\ve^2=e^{-\pi i/3}$, i.e. $i\ve^2 e^{\frac{\pi i(r+p-3)}{6k}}=1$,
and $(-1)^m e^{\pi i \frac{p}{k}m(m-r+1)}=1$. This yields the standard result
\beq\label{otr}
\gamma(\omega_1+\omega_2-\mu)\gamma(\mu)=e^{\pi i B_{2,2}(\mu;\omega_1,\omega_2)}.
\ee

Define now the normalized rarefied hyperbolic gamma function
\beq\label{gammakp}
\Gamma_M(\mu,m):=Z(m)e^{-\frac{\pi i}{2k}B_{2,2}(\mu;\omega_1,\omega_2)}\gamma_M(\mu,m),
\ee
where
\beq
Z(m)=\frac{e^{-\frac{\pi i}{4} (1+\frac{r+p-3}{3k})}}{\ve(-p,-s,k,-r)}
e^{\pi i \frac{(1-s)k-p}{2k}m(m-r+1)}.
\label{Z}\ee
The relation to the Dedekind sum $S(-d,c)$ (see formula (67.6) in \cite{rad} and Appendix A)
\beq
\varepsilon(a,b,c,d)=e^{\pi i S(-d,c)}e^{{\pi i\over 12c}(a+d)}
\ee
implies that  we can write $Z(m)$ also in the form
\beq\label{zmk}
Z(m)=e^{-\frac{\pi i}{4}(1-{1\over k})} e^{-\pi i S(r,k)}
e^{\pi i \frac{(1-s)k-p}{2k}m(m-r+1)}.
\ee
The multiplier $Z(m)$ guarantees the following simple reflection equation
\beq\label{reflekt}
\Gamma_M(\omega_1+\omega_2-\mu,r-1-m)\Gamma_M(\mu,m)= 1.
\ee
During its derivation one should use the identity $e^{\pi i (1-s)m(m-r+1)}=(-1)^{(1-s)m}$
following from the property $(r|s)=1$. Equality \eqref{reflekt} is equivalent to the relation
\beq
\Gamma_M(\pm\mu,\pm m):=\Gamma_M(\mu,m)\Gamma_M(-\mu,-m) =\frac{(-1)^{sm}}{4\sin \frac{\pi(\mu+m\omega_2)}{k\omega_2}
\sin \frac{\pi(-\mu+pm\omega_1)}{k\omega_1}}.
\ee

Note that $\Gamma_M(\mu,m)$ is an appropriate generalization of the function
\begin{equation}
\gamma^{(2)}(\mu;\omega):=e^{-\frac{1}{2}\pi i B_{2,2}(\mu;\omega_1,\omega_2)}\gamma(\mu;\omega),
\label{gamma2}\end{equation}
used in the previous considerations \cite{spi:conm} and \cite{SS}.
This gamma function is not periodic
\beq
\Gamma_M(\mu,m+k)= \xi\,\Gamma_M(\mu,m),\qquad \xi=e^{\pi i \frac{(1-s)k-p}{2}(k+2m-r+1)}.
\label{qusiper}\ee
One can check that $\xi^2=1$, i.e. the quasiperiodicity factor is a pure sign, $\xi=\pm1$.

Let us consider in more detail the rightmost exponential multiplier in \eqref{Z}.
From the expression \eqref{zmk} it follows that it is only the choice of this root of unity that
$\Gamma_M$ depends on the full set of discrete variables $k,p,r,s$, and the rest depends only on $p$ and $k$.
Indeed, under the first transformation $r\to r+nk$ and $s\to s-np$ keeping the
condition $pr+ks=1$ intact, one has
$$
e^{\pi i {(1-s)k-p\over 2k}m(m-r+1)}\to e^{\pi i {(1-s)k-p\over 2k}m(m-r+1)}
e^{{\pi i\over 2} mn(2p+2ks-k-1+pm-npk)}
$$
Under the second transformation $p\to p+nk$ and $s\to s-nr$ preserving $pr+ks=1$, one has
$$
e^{\pi i {(1-s)k-p\over 2k}m(m-r+1)}\to e^{\pi i {(1-s)k-p\over 2k}m(m-r+1)}
e^{{\pi i\over 2}n(r-1)m(m-r+1)}.
$$
One can check that both quasiperiodicity multipliers in these two relations are pure signs,
i.e. the $\Gamma_M$-dependence on other integer parameters in $M$ beyond $c$ and $d$ is minimal,
reduced to the sign choice.

To compute the residues in our integrals we will need the following limit
\beq\label{residue}
\lim_{\mu\to 0}\mu\,\Gamma_M(\mu,0)= \frac{\sqrt{\omega_1\omega_2}}{2\pi}\, k.
\ee
Indeed,
$$
\lim_{\mu\to 0}\mu\,\Gamma_M(\mu,0)= \frac{e^{-\frac{\pi i}{4} (1+\frac{r+p-3}{3k})}}{\ve(-p,-s,k,-r)}
e^{-\frac{\pi i}{2k}B_{2,2}(0;\omega_1,\omega_2)}
\lim_{\mu\to 0}\mu\,\gamma_M(\mu,0).
$$
Since
$$
\lim_{\mu\to 0}\mu\,\gamma_M(\mu,0)=-\frac{k\omega_2}{2\pi i}\frac{(\tilde q;\tilde q)_\infty}{(q;q)_\infty}
=-\frac{k\sqrt{-i\omega_1\omega_2}}{2\pi i}\, \ve(-p,-s,k,-r)\left( \frac{q}{\tilde q} \right)^{\frac{1}{24}},
$$
in combination we obtain \eqref{residue}.

This normalized function satisfies the equations
\beq
\frac{\Gamma_M(\mu+\omega_1,m+r)}{\Gamma_M(\mu,m)}
=(-1)^m e^{\pi i \frac{(r-1)(s-1)}{2}}
2\sin \frac{\pi(\mu +m\omega_2)}{k\omega_2},
\ee
where the sign factor in front of the sine-function represents
some quadratic character (note that $e^{\pi i (r-1)(s-1)}=1$),
and
\beq
\frac{\Gamma_M(\mu+\omega_2,m-1)}{\Gamma_M(\mu,m)}
=(-1)^{(s-1)m}e^{\pi i \frac{(r-1)(s-1)}{2}}
2\sin \frac{\pi(\mu -pm\omega_1)}{k\omega_1}.
\ee
Analogously,
$$
\frac{\Gamma_M(\mu-\omega_1,m-r)}{\Gamma_M(\mu,m)}
=\frac{(-1)^{m-r} e^{\pi i \frac{(r-1)(s-1)}{2}}}
{2\sin \frac{\pi(\mu+m\omega_2-(\omega_1+r\omega_2))}{k\omega_2}}
$$
and
$$
\frac{\Gamma_M(\mu-\omega_2,m+1)}{\Gamma_M(\mu,m)}
=\frac{(-1)^{(s-1)(m+1)}e^{\pi i \frac{(r-1)(s-1)}{2}} }
{2\sin \frac{\pi(\mu -pm\omega_1-(\omega_2+p\omega_1))}{k\omega_1}}.
$$
For $k=p=r=1$ one obtains the standard equations for $\gamma^{(2)}(u;\omega_1,\omega_2)$.

For further applications we need to know also the asymptotics of $\Gamma_M(\mu,m)$-function.
Assume that Im$(\omega_1/\omega_2)>0$.
With this condition the asymptotics of $\gamma(y;\omega_1,\omega_2)$ has the form
\bea \nonumber &&
\stackreb{\lim}{y\to \infty}\gamma(y;\omega_1,\omega_2)=1,
\quad {\rm for}\; {\rm arg}\;\omega_1<{\rm arg}\; y<{\rm arg}\;\omega_2+\pi,
\\ \nonumber &&
\stackreb{\lim}{y\to \infty}e^{-{\pi i}B_{2,2}(y;\omega_1,\omega_2)}\gamma(y;\omega_1,\omega_2)=1,
\quad {\rm for}\; {\rm arg}\;\omega_1-\pi<{\rm arg}\; y<{\rm arg}\;\omega_2.
\eea
Recalling relations (\ref{gampk}), (\ref{gammakp}), and (\ref{reflekt}), we obtain
\begin{eqnarray}\label{gmasymp1} &&
\lim_{\mu\to \infty}
\Gamma_M(\mu,m) \sim Z(m) e^{-\frac{\pi i}{2k}B_{2,2}(\mu;\omega_1,\omega_2)}, \;
\textrm{if}\; \arg \omega_1 < \arg \mu < \arg \omega_2 + \pi,
\\ \label{gmasymp2} &&
\lim_{\mu \to\infty}
\Gamma_M(\mu,m) \sim Z^{-1}(m) e^{\frac{\pi i}{2k}B_{2,2}(\mu;\omega_1,\omega_2)}, \;
\textrm{if}\; \arg \omega_1 -\pi< \arg \mu < \arg \omega_2.
\end{eqnarray}

\section{The rarefied hyperbolic beta integral}

Now we are going to construct a general univariate  rarefied hyperbolic beta integral
composed out of the function $\Gamma(\mu, m)$, which will be a hyperbolic analogue of the
elliptic beta integral \cite{spi:umn} for the general lens space. For proving its exact
evaluation formula we use an appropriate modification of the method suggested in \cite{spi:short}.

Let us take variables $\mu, a_j\in\CC$ and $m, n_j\in\Z+\nu$ with $j=1,\ldots, 6$ and $\nu=0,\frac{1}{2}$
and impose the balancing constraints
\beq
\sum_{j=1}^6 a_j=\omega_1+\omega_2,\qquad \sum_{j=1}^6n_j=r-1.
\label{balance}\ee
Define the function
\beq
\rho(\mu,m;\underline{a},\underline{n})=\frac{\prod_{j=1}^6\Gamma_M(a_j\pm \mu,n_j\pm m)}
{\Gamma_M(\pm 2\mu,\pm 2m)\prod_{1\leq \ell<j\leq 6} \Gamma_M(a_\ell+a_j,n_\ell+n_j)},
\label{rho}\ee
where we assume the notation
$$
\Gamma_M(a\pm \mu,n\pm m):=\Gamma_M(a+\mu,n+m)\Gamma_M(a- \mu,n- m).
$$
Let us resolve the balancing conditions by setting $a_6=\omega_1+\omega_2-A$ and
$n_6=r-1-N$, where $A=\sum_{j=1}^5a_j$ and $N=\sum_{j=1}^5n_j$.
Using the inversion relation for $\Gamma_M$-function we can write now
$$
\rho(\mu,m;\underline{a},\underline{n})=\frac{\prod_{j=1}^5\Gamma_M(a_j\pm \mu,n_j\pm m)}
{\Gamma_M(A\pm\mu, N\pm m)\Gamma_M(\pm 2\mu,\pm 2m)}
\frac{\prod_{j=1}^5 \Gamma_M(A-a_j, N-n_j)}
{\prod_{1\leq \ell<j\leq 5} \Gamma_M(a_\ell+a_j,n_\ell+n_j)}.
$$

Recurrence relations for the rarefied hyperbolic gamma function guarantee
validity of the following equation for the $\rho$-function
\bea \nonumber &&
\rho(\mu,m;a_1+\omega_1,a_2,\ldots,n_1+r,n_2,\ldots)-\rho(\mu,m;\underline{a},\underline{n})
\\ && \makebox[4em]{}
=g_1(\mu-\omega_1,m-r,\ldots;\underline{a},\underline{n})-g_1(\mu,m;\underline{a},\underline{n}),
\label{eqnrho1}\eea
where
\beq
\frac{g_1(\mu,m;\underline{a},\underline{n})}{\rho(\mu,m;\underline{a},\underline{n})}=
\frac{\prod_{j=1}^{5}\sin \frac{\pi(a_j+\mu +(n_j+m)\omega_2)}{k\omega_2} }
{\prod_{j=2}^{5}\sin \frac{\pi(a_1+a_j+(n_1+n_j)\omega_2)}{k\omega_2}}
\frac{\sin\frac{\pi(a_1+A+(n_1+N)\omega_2)}{k\omega_2}}
{\sin\frac{\pi (2\mu+2m\omega_2)}{k\omega_2}\sin\frac{\pi(A+\mu+(m+N)\omega_2)}{k\omega_2}}.
\label{g1/rho}\ee
One has the ratios
\bea \nonumber &&
\frac{\rho(\mu,m;a_1+\omega_1,\ldots, n_1+r,\ldots)}{\rho(\mu,m;\underline{a},\underline{n})}=
\frac{\sin \frac{\pi (a_1\pm \mu+(n_1\pm m)\omega_2)}{k\omega_2}}
{\sin \frac{\pi (A\pm \mu+(N\pm m)\omega_2)}{k\omega_2}}
\prod_{j=2}^5 \frac{\sin \frac{\pi (A-a_j+(N-n_j)\omega_2)}{k\omega_2}}
 {\sin \frac{\pi (a_1+a_j+(n_1+n_j)\omega_2)}{k\omega_2}},
\\ \nonumber &&
\frac{\rho(\mu-\omega_1,m-r;\underline{a},\underline{n})}{\rho(\mu,m;\underline{a},\underline{n})}=
\prod_{j=1}^5\frac{\sin \frac{\pi (a_j-\mu+(n_j-m)\omega_2)}{k\omega_2}}
{\sin \frac{\pi (a_j+\mu+(n_j+m)\omega_2- (\omega_1+r\omega_2))}{k\omega_2}}
\\ && \makebox[4em]{}
\times
 \frac{\sin \frac{\pi (A+\mu+(N+m)\omega_2- (\omega_1+r\omega_2))}{k\omega_2}}
 {\sin \frac{\pi (A-\mu+(N-m)\omega_2)}{k\omega_2}}
\frac{\sin \frac{2\pi (\mu+m\omega_2- \omega_1-r\omega_2)}{k\omega_2}}
{\sin{\frac{2\pi (\mu+m\omega_2)}{k\omega_2}}}.
 \nonumber \eea
Therefore, dividing equation (\ref{eqnrho1}) by $\rho(\mu,m;\underline{a},\underline{n})$ we obtain
the following trigonometric identity
\bea \nonumber  &&
\frac{\sin \frac{\pi (a_1\pm \mu+(n_1\pm m)\omega_2)}{k\omega_2}}
{\sin \frac{\pi (A\pm \mu+(N\pm m)\omega_2)}{k\omega_2}}
\prod_{j=2}^5 \frac{\sin \frac{\pi (A-a_j+(N-n_j)\omega_2)}{k\omega_2}}
 {\sin \frac{\pi (a_1+a_j+(n_1+n_j)\omega_2)}{k\omega_2}} -1
\\ && \makebox[2em]{}
= \frac{\sin \frac{\pi (A+a_1+(N+n_1)\omega_2)}{k\omega_2}}
{\sin{\frac{2\pi (\mu+m\omega_2)}{k\omega_2}}
\prod_{j=2}^{5}\sin \frac{\pi(a_1+a_j+(n_1+n_j)\omega_2)}{k\omega_2}  }
\\ && \makebox[2em]{}  \times
\left(\frac{\prod_{j=1}^5\sin \frac{\pi (a_j-\mu+(n_j-m)\omega_2)}{k\omega_2}}
{\sin \frac{\pi (A-\mu+(N-m)\omega_2)}{k\omega_2}}
-\frac{\prod_{j=1}^{5}\sin \frac{\pi(a_j+\mu +(n_j+m)\omega_2)}{k\omega_2} }
{\sin \frac{\pi (A+ \mu+(N+ m)\omega_2)}{k\omega_2}}\right),
\nonumber \eea
or
\bea \nonumber  &&
\frac{1-t_1z^{\pm1}}{1-Tz^{\pm1}}
\prod_{j=2}^5 \frac{1-Tt_j^{-1}}{1-t_1t_j} -1
\\ && \makebox[2em]{}
= \frac{t_1 (1-t_1T)}{z(1-z^2)\prod_{j=2}^{5}(1-t_1t_j)}
\left(z^4\frac{\prod_{j=1}^{5}(1-t_jz^{-1})}{1-Tz^{-1}}-\frac{\prod_{j=1}^{5}(1-t_jz)}{1-Tz}\right),\quad\quad
\label{base_eq} \eea
where
$$
z=e^{2\pi i\frac{\mu+m\omega_2}{k\omega_2}}, \quad t_j=e^{2\pi i\frac{a_j+n_j\omega_2}{k\omega_2}}, \quad
T=e^{2\pi i\frac{A+N\omega_2}{k\omega_2}}.
$$
This relation is precisely the $e^{2\pi i\tau}\to 0$ case of the elliptic function identity established in \cite{spi:short}
(see also \cite{spi:rare}).

Analogously, one proves the equation
\bea \nonumber &&
\rho(\mu,m;a_1+\omega_2,a_2,\ldots,n_1-1,n_2,\ldots)-\rho(\mu,m;\underline{a},\underline{n})
\\ && \makebox[4em]{}
=g_2(\mu-\omega_2,m+1,\ldots;\underline{a},\underline{n})-g_2(\mu,m;\underline{a},\underline{n}),
\label{eqnrho2}\eea
where $g_2/\rho$-function is obtained from $g_1/\rho$ in (\ref{g1/rho}) by the replacement of
$\omega_2\to -p\omega_1$
in the numerators of the arguments of sin-functions and $\omega_2\to \omega_1$ in the denominators.
This results again in the equation (\ref{base_eq}) with
$$
z=e^{2\pi i\frac{\mu-mp\omega_1}{k\omega_1}}, \quad t_j=e^{2\pi i\frac{a_j-n_jp\omega_1}{k\omega_1}}, \quad
T=e^{2\pi i\frac{A-Np\omega_1}{k\omega_1}}.
$$

It can be checked that the functions $\rho$ and $g_{1,2}$ are $k$-periodic
\beq
f(\mu,m+k;\underline{a},\underline{n})=f(\mu,m;\underline{a},\ldots, n_j+k,\ldots)=f(\mu,m;\underline{a},\underline{n}), \quad j=1,2\ldots,5.
\label{period}\ee
Suppose that $\textrm{Re}(\omega_j)>0,\,\textrm{Im}(\omega_j)>0$.
Then the poles of $\Gamma_M(\mu,m)$
are located in the left lower quarter of the complex plane, and zeros lie in the right upper
quarter. Denote now
\beq\label{I}
I(\underline{a},\underline{n}):=\sum_{m\in\Z_k+\nu}\int_{-i\infty}^{i\infty}\rho(\mu,m;\underline{a},\underline{n})\, d\mu,
\ee
where $\Z_k=\{0,1,\ldots,k-1\}$ and $\nu=0, \frac{1}{2}$.
We impose also the restrictions $\textrm{Re}(a_\ell)>0,$
which ensure that all pole arrays of the integrand lie either to the right or to the left of the imaginary axis.

To prove the convergence of this integral, recall the asymptotics of the $\Gamma_M$ function (\ref{gmasymp1}), (\ref{gmasymp2}).
Substitute these relations into the kernel $\rho$ for the limit $\mu = +i\lambda,\, \lambda\to +\infty$.
This yields the asymptotics
\beq
\lim_{\lambda\to+ \infty} \rho(\mu,m;\underline{a},\underline{n})\propto
e^{-\frac{6\pi\lambda}{k}\left(\frac{1}{\omega_1}+\frac{1}{\omega_2}\right)}
\label{as}\ee
Since we took $\textrm{Re}(\omega_{1,2}) > 0$ we see that the kernel vanishes exponentially
fast and the integral converges. Similar exponential fallout takes place in the limit
$\lambda \to - \infty$, i.e. the kernel vanishes sufficiently fast even after the shift
of $\mu$ by any constant, which will be necessary below.

Actually the contour of integration can be chosen in a substantially more general
form -- it is only necessary to demand convergence of the integral
in the indicated domain of period values $\omega_{1,2}$. In particular, choosing appropriately this
contour one can relax taken restrictions on the $\omega_{1,2}$-variables.

Using equalities \eqref{eqnrho1} and \eqref{eqnrho2}, for the function \eqref{I} we obtain the equations
\bea\nonumber &&
I(a_1+\omega_{1},a_2,\ldots,n_1+r,n_2\ldots)-I(\underline{a},\underline{n})
\\ \nonumber && \makebox[2em]{}
=
\sum_{m\in\Z_k+\nu}\int_{-i\infty}^{i\infty}(g_{1}(\mu-\omega_{1},m-r;\underline{a},\underline{n})-g_{1}(\mu,m;\underline{a},\underline{n}))\, d\mu,
\\ \nonumber &&
I(a_1+\omega_{2},a_2,\ldots,n_1-1,n_2\ldots)-I(\underline{a},\underline{n})
\\ \nonumber && \makebox[2em]{}
=
\sum_{m\in\Z_k+\nu}\int_{-i\infty}^{i\infty}(g_{2}(\mu-\omega_{2},m+1;\underline{a},\underline{n})-g_{2}(\mu,m;\underline{a},\underline{n}))\, d\mu.
\nonumber\eea
Because of the $k$-periodicity of the functions $g_{1,2}$ in the variable $m$, we can replace $m-r$ by $m$
in the first equation, and $m+1$ by $m$ in the second.

Let us impose such constraints on the parameters $a_\ell$ that there will not be poles
in the vertical stripes of $\mu$ bounded by the points $-\textrm{Re}(\omega_1)$ and $0$, as well as
by the points $-\textrm{Re}(\omega_2)$ and $0$.
The $\mu$-dependent part of the $g_1(\mu,m;\underline{a},\underline{n})$-function has the form
$$
\frac{\prod_{\ell=1}^5\Gamma_M(a_\ell-\mu,n_\ell- m)\Gamma_M(a_\ell+\mu+\omega_1,n_\ell+m+r)}
{\Gamma_M(A-\mu, N- m)\Gamma_M(A+\mu+\omega_1,N+m+r)}\sin \frac{\pi(-2\mu+2pm\omega_1)}{k\omega_1}.
$$
The poles of this function in $\mu$ are located at the points
$$
a_\ell-\mu,\, a_\ell+\mu+\omega_1= -j\omega_1-(kn+n_l-m+jr)\omega_2,\quad \ell=1,\ldots, 5,
$$
where $j,\, kn+n_\ell-m+jr \geq 0$,  and
$$
A-\mu,\, A+\mu+\omega_1 = (p(N+m+r+j+1)+kn)\omega_1 +(j+1)\omega_2,
$$
where $j+1,\, p(N+m+r+j+1)+kn > 0$.
One can see that the first set of points does not have representatives in the vertical strip
bounded by the points $-\textrm{Re}(\omega_1)$ and $0$. The second set points do not enter this
region, if $\textrm{Re}(\omega_2-A)>0$. Imposing the latter condition we find
\beq
I(a_1+\omega_{1},a_2,\ldots,n_1+r,n_2\ldots)=I(\underline{a},\underline{n}).
\ee
After imposing the constraint  $\textrm{Re}(\omega_1-A)>0$, the second equation for the
$\rho$-function yields the equality
\beq
I(a_1+\omega_{2},a_2,\ldots,n_1-1,n_2\ldots)=I(\underline{a},\underline{n}).
\ee
Repeating these relations $k$ times and using the $k$-periodicity we find
\beq
I(a_1+k\omega_{1},a_2,\ldots,\underline{n})=I(a_1+k\omega_{2},a_2,\ldots,\underline{n})
=I(\underline{a},\underline{n}).
\ee
Note that the function $\Gamma_M(\mu,m)$ is well defined for $\omega_{1,2}>0$ (i.e., when $|q|=1$).
But for incommensurate real periods $\omega_1$ and $\omega_2$ the derived relations show that $I(\underline{a},\underline{n})$
is a constant independent of $a_\ell$ and $n_\ell$. Let us compute this constant
using the residue calculus.

Let us take the limit when two pairs of poles of a particular integral entering $I(\underline{a},\underline{n})$
start to pinch the contour of integration. To find it, we indicate the poles of the integrand
coming from the $\Gamma_M$-function poles:
$$
a_\ell\pm\mu=-j\omega_1-j'\omega_2,\quad j':=n_\ell\pm m +rj+kn\geq 0,
\quad  j\in\Z_{\geq0},\; n\in\Z, \; \ell=1,\ldots,6.
$$
Now we fix the values of $n_\ell$ as follows
\beq
n_1=n_2=n_3=\nu,\quad n_4=n_5=-\nu, \quad N=\sum_{\ell=1}^5n_\ell=\nu,\quad \nu=0,\frac{1}{2},
\ee
and take the limit
$
a_1+a_4\to 0.
$
As a result, two pairs of poles pinch the integration contour;
the first pair being $\mu =-a_1, a_4$, which satisfies necessary conditions for $m=\nu$,
and the second one $\mu=a_1, -a_4$ emerging for $m=0$, if $\nu=0$, and $m=k-\nu$, if $\nu=1/2$.

Deform now the contour of integration to the left, pick up residues of the poles
at $\mu=-a_1,-a_4$ and take the limit $a_1+a_4\to 0$. Applying the Cauchy theorem we find
\beq
I(\underline{a},\underline{n})=2\pi i(\lim_{\mu\to -a_4}(\mu+a_4)\rho(\mu,\nu;\underline{a},\underline{n})+
\lim_{\mu\to -a_1}(\mu+a_1)\rho(\mu,k-\nu;\underline{a},\underline{n}))=2i\sqrt{\omega_1\omega_2}k.
\ee
E.g.,
\bea\nonumber
&&\lim_{\mu\to -a_1}(\mu + a_1)\rho(\mu, k-\nu; \underline{a} , \underline{n}) =
\lim_{\mu\to -a_1}(\mu + a_1)\Gamma_M(a_1 +\mu, 0)\\ \nonumber
&&\times\frac{\prod_{\ell =2}^3\Gamma_M(a_\ell-a_1, 0)\Gamma_M(a_\ell + a_1, 2\nu)
\prod_{\ell =4}^5 \Gamma_M(a_\ell-a_1,-2\nu)\, \Gamma_M(a_\ell + a_1,0)}
{\Gamma_M(A -a_1, 0)\Gamma_M(A + a_1, 2\nu)\Gamma_M(-2a_1,-2\nu)}\\ \nonumber
&&\times \frac{\prod_{\ell =1}^3\Gamma_M(A -a_\ell, 0)\prod_{\ell =4}^5\Gamma_M(A-a_\ell,2\nu)}
{\prod_{1\leq \ell <k\leq3}\Gamma_M(a_\ell + a_k, 2\nu)
\prod_{\ell =1}^3\prod_{k=4}^5\Gamma_M(a_\ell + a_k, 0)\,\Gamma_M(a_4 + a_5,-2\nu)}
=\frac{\sqrt{\omega_1\omega_2}}{2\pi}\,k,
\eea
where $A = a_2 + a_3 + a_5$. Thus we have proved the following theorem.

{\bf Theorem.} Let $a_j\in\CC,\, \textrm{Re}(a_j) > 0,\, j = 1,\ldots,6,$
and $\textrm{Re}(\omega_{1,2}) > 0$.
Also take $n_j\in \Z +\nu,\,  j = 1,\ldots,6$, $\nu = 0,\frac{1}{2}$
and impose the following balancing condition
\beq
\sum_{j=1}^6a_j = \omega_1 + \omega_2,\qquad \sum_{j=1}^6 n _j=r-1.
\label{balancing}\ee
Then the following identity holds:
\bea && \makebox[-3em]{}
\sum_{m\in\Z_k+\nu}\int_{-i\infty}^{i\infty}
\frac{\prod_{j=1}^6\Gamma_M(a_j\pm \mu,n_j\pm m)}
{\Gamma_M(\pm 2\mu,\pm 2m)} \, \frac{d\mu}{2ik\sqrt{\omega_1\omega_2}}
= \prod_{1\leq \ell<j\leq 6} \Gamma_M(a_\ell+a_j,n_\ell+n_j),
\label{integral}\eea
where $\Z_k = \{0, 1,\ldots,k -1\}$.

The constraints $\textrm{Re}(\omega_{1,2}-A)>0$ (or $\textrm{Re}(a_6-\omega_{1,2})>0$)
used in the proof of the theorem, as well as restrictions on $\omega_{1,2}$,
are lifted by the analytical continuation.
Equivalently, relation \eqref{integral} can be rewritten in terms of the $\gamma_M$-function
\bea\nonumber &&
\sum_{m\in\Z_k+\nu}\int_{-i\infty}^{i\infty}
e^{-\frac{2\pi i p}{k}\,(m^2-\nu^2)}
e^{-\frac{2\pi i}{k\omega_1\omega_2}\,\mu^2} \,
\frac{\prod_{j=1}^6\gamma_M(a_j\pm \mu,n_j\pm m)}
{\gamma_M(\pm 2\mu,\pm 2m)} \, \frac{d\mu}{2ik\sqrt{\omega_1\omega_2}}
\\  \nonumber && \makebox[2em]{}
= e^{-2\pi i \frac{(1-s)k-p}{k}\left(  N+\nu^2 \right)}
e^{\pi i\left( \frac{1}{\omega_1\omega_2k}(\frac{7}{12}(\omega_1+\omega_2)^2-\sum_{j=1}^6a_j^2)
-\frac{5}{4}(1-\frac{2}{3k})-5S(r,k)\right)}
\\ && \makebox[2em]{} \makebox[2em]{} \times
\prod_{1\leq  \ell < j \leq 6} \gamma_M(a_\ell+a_j,n_\ell+n_j),
 \label{integral2} \eea
where $N=\sum_{1\leq  \ell < j \leq 6}n_\ell n_j$.
Under the shift $n_j\to n_j+k$ for some fixed
$j$ one has $N\to N+k\sum_{\ell=1,\neq j}^6 n_\ell$. Therefore, for $\nu=0$
 both sides of the equality \eqref{integral2} are invariant with respect to such shifts.
Correspondingly, in this case we can set $n_j\in\Z_k$ and take a slightly more general
discrete balancing condition $\sum_{j=1}^6 n _j=r-1\mod k.$
However, if $\nu=1/2$, then $N$ is shifted by the half-integer and an analogous statement
will not be true. In this case we can replace the multiplier $Z(m)$ in \eqref{gammakp}
by the exponential of a cubic polynomial of $m$ which, in difference from \eqref{qusiper},
will guarantee the periodicity $\Gamma_M(\mu,m+k)= \Gamma_M(\mu,m)$. It will
produce in the right-hand side of \eqref{integral2} a different $a_j$-independent
multiplier, which will be invariant under the shifts $n_\ell\to n_\ell+k$
for all $\nu$ and which will coincide with the one given above for the reduced
balancing condition \eqref{balancing}.

Since we use the general modular transformation for the Dedekind $\eta$-function
and Jacobi theta-function, the theory of rarefied hyperbolic beta integrals can be
considered as a complement to the theory of Jacobi forms \cite{EZ}, because the kernels
of integrals are composed of ``one-halves'' of the meromorphic Jacobi forms
(in the sense of the number of divisor points).

The key identity (\ref{integral}) can be rewritten in the form of the star-triangle relation,
which leads to a new solvable model of $2d$ lattice spin system, and as a consequence to new
solutions of the Yang-Baxter equation similar to \cite{BMS,BT06,CD14,CS,kashaev,kashaevYBE,VF}.
These and some other applications, as well as the problem of elliptic
generalization of the equality \eqref{integral} will be considered in a separate work.

\section{Limiting relations}

For simplicity we consider further only the $\nu=0$ case.
Let us reparametrize $a_j, n_j$ in the identity (\ref{integral}) in the following asymmetric way
\bea
&& a_j= f_j+i\xi, \quad a_{j+3}= g_{j}-i\xi, \quad l_j:=n_{j+3}, \quad j=1,2,3.
\eea
Then the balancing condition takes the form
\beq\label{bal}
\sum_{j=1}^3(f_j+g_j)=\omega_1+\omega_2, \quad \sum_{j=1}^3(n_j+l_j)=r-1.
\ee

Now we shift in \eqref{integral} the integration variable $\mu\to \mu-i\xi$ and take the limit
$\xi\to -\infty$ using the asymptotics of $\Gamma_M(\mu,m)$.
Since the integrand is an even function (in fact the parity
transformation reshuffles the separate terms keeping the sum intact), one can write
\bea \nonumber
&&
2\int_{0}^{i\infty}\sum_{m=0}^{r-1}\left[{\prod_{j=1}^3\Gamma_M(\mu+f_j+i\xi,n_j+m)
\Gamma_M(\mu+g_j-i\xi,l_{j}+m)
\over \Gamma_M(2\mu,2m)\Gamma_M(-2\mu,-
2m)}\right.
\\&& \makebox[2em]{}
\times \left.\prod_{j=1}^3\Gamma_M(-\mu+f_j+i\xi,n_j-m)\Gamma_M(-\mu+g_j-i\xi,l_{j}-m)\right]
{d\mu\over 2ik\sqrt{\omega_1\omega_2}}
\\ \nonumber &&  \makebox[2em]{}
=2\int_{i\xi}^{i\infty}\sum_{m=0}^{r-1}\prod_{j=1}^3\Gamma_M(\mu+f_j,n_j+m)
\Gamma_M(-\mu+g_j,l_j-m)e^{-{\pi i\over 2k}\sigma_1}
{d\mu\over 2ik\sqrt{\omega_1\omega_2}},
\eea
where in the limit $\xi\to -\infty$
\bea && \makebox[-2em]{}
\sigma_1=\sum_{j=1}^3\left[B_{2,2}(\mu+g_j-2i\xi)
-B_{2,2}(-\mu+f_j+2i\xi)\right]
-B_{2,2}(2\mu-2i\xi)+ B_{2,2}(-2\mu+2i\xi)
\nonumber \\ && \makebox[-2em]{}
-((1-s)k-p)\left(\sum_{j=1}^3 \left[(l_{j}+m)^2-(r-1)(l_{j}+m)-(n_{j}-m)^2+(r-1)(n_{j}-m)\right]
+4m(r-1)\right).
\nonumber\eea
On the right-hand side of equality (\ref{integral}) we have
\bea \nonumber  &&
\prod_{\ell,j=1}^3\Gamma_M(f_\ell+g_j,n_\ell+l_j)
e^{-{\pi i\over 2k}\sigma_2}, \quad
\sigma_2=\sum_{1\leq i<j\leq 3}\left[B_{2,2}(g_i+g_j-2i\xi)
-B_{2,2}(f_i+f_j+2i\xi)\right]
\\ && \makebox[-1em]{}
-((1-s)k-p)\sum_{1\leq \ell<j\leq 3}\left[(l_\ell+l_j)^2
-(r-1)(l_\ell+l_j)-(n_\ell+n_j)^2+(r-1)(n_\ell+n_j)\right].
\nonumber \eea
Similar to the considerations of \cite{spi:conm} for $k=1$ case,
it can be checked that all $B_{2,2}$-terms appearing on the left- and right-hand sides
cancel each other. Taking care about the rest yields:
\bea  \makebox[-1em]{}
\int_{-i\infty}^{i\infty}\sum_{m=0}^{r-1}\prod_{j=1}^3\Gamma_M(\mu+f_j,n_j+m)
\Gamma_M(-\mu+g_j,l_j-m){d\mu\over ik\sqrt{\omega_1\omega_2}}
=\prod_{\ell,j=1}^3 \Gamma_M(f_\ell+g_j,n_\ell+l_j).
\label{namer}\eea

Let us compare this result with the analogous formula in paper \cite{SS} for the parafermionic
hyperbolic gamma function corresponding to the choice $p=r=1$ and $s=0$.
That formula in \cite{SS} contains an additional sign factor in the integral.
To see how it appears we should compare the definition of $\Gamma_M$ (\ref{gammakp}) for $p=1$
with the definition in \cite{SS},
\bea \nonumber
&&\Lambda(y, m;\omega_1,\omega_2)=\prod_{k=0}^{m-1}\gamma^{(2)}
\left({y\over r}+\omega_2\left(1-{m\over r}\right)+(\omega_1+\omega_2){k\over r};\omega_1,\omega_2\right)
\\ \nonumber && \makebox[4em]{} \times
\prod_{k=0}^{r-m-1}\gamma^{(2)}
\left({y\over r}+{m\over r}\omega_1+(\omega_1+\omega_2){k\over r};\omega_1,\omega_2\right)\, ,
\eea
Remembering that \cite{rad2}
$S(1,k)=-{1\over 4}+{1\over 6k}+{k\over 12}$
and using (\ref{zmk}) one can write
$$
\Gamma_{\tiny\left(
\begin{array}{cc}
-1 & 0  \\
k & -1
\end{array} \right)}(\mu,m)=e^{{\pi i\over 12}({1\over k}-k)}e^{\pi i{k-1\over 2k}m^2}e^{-{\pi i\over 2k}B_{2,2}(\mu;\omega_1,\omega_2)}
\gamma_{\tiny\left(
\begin{array}{cc}
-1 & 0  \\
k & -1
\end{array} \right)}(\mu,m).
$$
On the other hand, using the definition \eqref{gamma2} we obtain
$$
\Lambda(\mu, m;\omega_1,\omega_2)=e^{{\pi i\over 12}({1\over k}-k)}e^{\pi i{km-m^2\over 2k}}e^{-{\pi i\over 2k}B_{2,2}(\mu;\omega_1,\omega_2)}
\gamma_{\tiny\left(
\begin{array}{cc}
-1 & 0  \\
k & -1
\end{array} \right)}(\mu,m).
$$
Finally, we come to the relation
\beq
\Gamma_{\tiny\left(
\begin{array}{cc}
-1 & 0  \\
k & -1
\end{array} \right)}(\mu,m)=e^{{\pi i\over 2}(m^2-m)}\Lambda(\mu, m;\omega_1,\omega_2)
\ee
and it is this sign difference between $\Gamma_M$ and $\Lambda$ that gives rise to the sign
factor in the integral considered in \cite{SS}.

Now we can obtain three more integral relations taking various limits of the parameters $f_\ell$ and $g_\ell$.
Resolving the balancing condition (\ref{bal}) for $g_3$ and taking the limit $f_3\to i\infty$ in the
(\ref{namer}) we obtain:
\bea \nonumber
&&\int_{-i\infty}^{i\infty}\sum_{m=0}^{k-1}e^{{\pi i m\over k}(p-k(1-s))\left(n_1+n_2+l_1+l_2\right)}
e^{\left({\pi i\over \omega_1\omega_2 k}\left[ y(f_1+f_2+g_1+g_2)+f_1f_2-g_1g_2\right]\right)}
\Gamma_M(y+f_1,n_1+m)
\\  \label{namer33} && \makebox[2em]{} \times
\Gamma_M(y+f_2,n_2+m)
\Gamma_M(-y+g_1,l_1-m)
\Gamma_M(-y+g_2,l_2-m)
{dy\over i\sqrt{\omega_1\omega_2}}
\\ \nonumber && \makebox[1em]{}
=ke^{{\pi i\over k}\left[(p-k(1-s))\left(l_1l_2-n_1n_2\right)\right]}
\Gamma_M(\omega_1+\omega_2-f_1-f_2-g_1-g_2,r-1-n_1-l_1-n_2-l_2)
\\ \nonumber &&  \makebox[2em]{}  \times
\Gamma_M(f_1+g_1,n_1+l_1)\Gamma_M(f_1+g_2,n_1+l_2)
\Gamma_M(f_2+g_1,n_2+l_1)\Gamma_M(f_2+g_2,n_2+l_2).
\eea
Further on, taking in (\ref{namer33}) the limit $f_2\to -i\infty$ and $g_2\to i\infty$ with $f_2+g_2=\alpha$
kept fixed, we obtain
\bea\label{namer2} &&
\int_{-i\infty}^{i\infty}\sum_{m=0}^{k-1}e^{{\pi i m\over k}\left[(p-k(1-s))\left(2N+l+1-r\right)\right]}
e^{{\pi i\over \omega_1\omega_2 k}\left[g\left({Q\over 2}-\alpha\right) -{g^2\over 2}+y(2\alpha+g-Q)
\right]}
\\ \nonumber  &&  \makebox[4em]{} \times  \Gamma_M(y,m)\Gamma_M(-y+g,l-m)
{dy\over i\sqrt{\omega_1\omega_2}}
\\ \nonumber && \makebox[1em]{}
=ke^{\left[{\pi i\over 2k}(p-k(1-s))l(1-r+(l+2N))\right]}
\Gamma_M(Q-\alpha-g,r-1-N-l)
\Gamma_M(\alpha,N)\Gamma_M(g,l).
\eea
Here $N:=n_2+l_2$, $Q:=\omega_1+\omega_2$. We also denoted $g_1=g,\, l_1=l$ and set $f_1=0$, $n_1=0$ since
these variables can be restored by the shifts $y\to y+f_1, \, g\to g+f_1$ and $m\to m+n_1,\, l\to l+n_1$.

Finally, taking in equality (\ref{namer2}) the limit $g\to -i\infty$, we obtain
\bea\label{namer3} &&
e^{\frac{\pi i}{4}(1-{1\over k})} e^{\pi i S(r,k)}e^{-{\pi i\over \omega_1\omega_2 k}{\omega_1^2+\omega_2^2\over 24}}
\int_{-i\infty}^{i\infty}\sum_{m=0}^{k-1}e^{{\pi i m\over k}(p-k(1-s))\left(2N+{1\over 2}(1-r)\right)}
e^{{\pi i m^2\over 2k}(p-k(1-s))}
\\ \nonumber && \makebox[-2em]{} \times
e^{{\pi i\over \omega_1\omega_2 k}\left[{y^2\over 2}-2y\left({Q\over 4}-\alpha\right)\right]}
 \Gamma_M(y,m)
{dy\over i\sqrt{\omega_1\omega_2}}
=ke^{{\pi i\over 2k}(p-k(1-s))\left(N(r-1)-N^2\right)}e^{-{\pi i\over 2\omega_1\omega_2k} \left({Q\over 2}-\alpha\right)^2}
\Gamma_M(\alpha,N).
\eea

To compare these integrals with those computed by Dimofte in \cite{dimofte}, note that the function
${\mathcal Z}^{(k,p)}_b(y,m)$ used in \cite{dimofte} is related to $\Gamma_M(y,m)$
after setting $\omega_1=b^{-1}, \omega_2=b$, by the relation
\beq\label{gzm}
\Gamma_M(y,m)=Z(m)^{-1}e^{{\pi i\over 2k}B_{2,2}(y;b,b^{-1})}{\mathcal Z}^{(k,p)}_b(iy,m),
\ee
which can be obtained from formula (\ref{gamdim}) in Appendix B. Note that here $b$ is
the $q$-deformation parameter and it should not be mixed with the integer $b$-variable entering
the description of modular group transformation \eqref{PSL}.

Inserting (\ref{gzm}) in (\ref{namer2}) we obtain
\bea \nonumber &&
e^{-\frac{\pi i}{4}(1-{1\over k})} e^{-\pi i S(r,k)}e^{{\pi i\over  k}{2Q^2-1\over 12}}
\int_{-\infty}^{\infty}\sum_{m=0}^{k-1}(-1)^{(1-s)m}e^{{\pi i p\over k}\left(m^2+2Nm+(1-r)m\right)}
\\  \label{namer22} && \makebox[2em]{} \times
e^{{\pi i\over  k}\left[-y^2+y(-2\alpha+iQ)+\left({iQ\over 2}-\alpha-g\right)^2\right]}
{\mathcal Z}^{(k,p)}_b(y,m){\mathcal Z}^{(k,p)}_b(-y+g,l-m) dy
\\ \nonumber && \makebox[-2em]{}
=ke^{{\pi i\over k}(p-k(1-s))\left[\left({r-1\over 2}-N-l\right)^2-{(r-1)^2\over 4}\right]}
{\mathcal Z}^{(k,p)}_b(iQ-\alpha-g,r-1-N-l)
{\mathcal Z}^{(k,p)}_b(\alpha,N){\mathcal Z}^{(k,p)}_b(g,l).
\eea
This integral relation has been suggested first and verified numerically for $p=r=1$ and $s=0$ in \cite{Imamura:2012rq,Imamura:2013qxa}.
Then it has been proved in  \cite{dimofte} for the case
of even $s$ and odd $p$ and $r$, and our result confirms the corresponding computations.
Note that in \cite{dimofte} this relation is written in terms of the function
$\kappa(k,r)=3(k-1)^2-{12\over k}\sum_{j=0}^{k-1}j\left(rj+{r-1\over 2}\;{\rm mod}\; k\right)$.
To see that both formulae indeed coincide one should take into account that for odd $r$
the function $\kappa(k,r)$ satisfies the relation
$$
e^{-{\pi i\over 6k} \kappa(k,r)}=e^{\pi i F\left({r-1\over 2}\right)}=e^{{\pi i\over 2}\left(1-{1\over k}\right)}e^{2\pi i S(r,k)}e^{{\pi i p\over k}{r^2-1\over 4}}(-1)^{r-1\over 2} e^{-{\pi i\over k}{r-1\over 2}},
$$
where the function $F(m)$ is described in Appendix B.

Now inserting (\ref{gzm}) in (\ref{namer3}) we obtain
\bea\label{namer34}
&&e^{\frac{\pi i}{4}(1-{1\over k})} e^{\pi i S(r,k)}e^{{\pi i\over  k}{Q^2+1\over 12}}
\int_{-\infty}^{\infty}\sum_{m=0}^{k-1}(-1)^{(1-s)m}e^{{\pi i p \over k}\left(m^2+ 2Nm+m(1-r)\right)}
\\ \nonumber &&  \makebox[2em]{}  \times
e^{{\pi i\over  k}\left[-y^2+2y\left(i{Q\over 2}-\alpha\right)\right]}
 {\mathcal Z}^{(k,p)}_b(y,m)dy=k{\mathcal Z}^{(k,p)}_b(\alpha,N).
\eea
For $p=r=1$ and $s=0$ this relation reduces to the equality considered in  \cite{dimofte}.
The identity (\ref{namer34}) expresses a $3d$ mirror symmetry between the theory of free chiral
field and the $U(1)$ gauge theory with a chiral field with the $1/2$ Chern-Simons coupling
on the lens space $L(k,k-p)_\tau$ \cite{Dimofte:2011ju}. The term $e^{{\pi i p m^2\over k}}$
gives a contribution of the flat connection with the holonomy $m$ to the Chern-Simons action
in agreement with \cite{Griguolo:2006kp,hansen}.

As mentioned above, general modular quantum dilogarithm can be written as a product
of a number of hyperbolic gamma functions with arguments lying on a specific
lattice of points  \cite{dimofte}. It is interesting to note that precisely
the same special lattice of points emerged first in  \cite{Bonelli} in the products of
so-called $\Upsilon$-functions that were used for writing structure constants
of an $2d$ conformal field theory related through the AGT correspondence to ${\mathcal N}=2$
four-dimensional gauge theory on $\mathbb{C}^2/\Gamma_{c,d}$, where $\Gamma_{c,d}\subset U(2)$
is a finite group acting on local coordinates according to formula \eqref{lens_space}.
The lattice appearing in a particular case $d=-1$ was introduced
earlier in the paper \cite{BFL} for description of structure constants in the quantum
Liouville field theory interacting with parafermions (para Liouville field theory).
It is expected that the general rarefied hyperbolic gamma function will play a similar
important role in the mentioned $2d$ conformal field theory for description of the
fusion matrix and boundary correlation functions.

\smallskip

The authors are indebted to A. B. Kalmynin and R. M. Kashaev for a discussion of obtained results.
This work is partially supported by Laboratory of Mirror Symmetry NRU HSE, RF government grant,
ag. no. 14.641.31.0001.

\appendix
\section{The Dedekind sum}

The Dedekind function is defined as the following sum \cite{rad2}
\beq\label{Ded}
S(r,k)=\sum_{\delta=1}^{k-1}{\delta\over k}
\left({r\delta\over k}-\left[{r\delta\over k}\right]-{1\over 2}\right),
\ee
where $[x]$ is the integer part of $x\in\mathbb{R}$.
Its key properties are
\bea \nonumber &&  \makebox[2em]{}
S(-r,k)=-S(r,k),\qquad S(r,k)=S(p,k),
\\ \label{DedProp}  &&
S(r,k)={1\over 4k}\sum_{m=1}^{k-1}\cot{\pi m\over k}\cot{\pi r m\over k}, \quad
S(r+k,k)=S(r,k).
\eea

\section{Dimofte's notations}

The original modular quantum dilogarithm \cite{Fad94,Fad95} can be written in the form \cite{BMS}
\beq
\phi(z)=\exp\left({1\over 4}\int_{\mathbb{R}+i\epsilon}{dw\over w}{e^{-2\pi i z w}\over \sinh(\pi b w)
\sinh(\pi b^{-1} w)}\right).
\ee
In \cite{dimofte} the function ${\mathcal Z}^{(1,1)}_b(z,0)$ is defined as
\beq
{\mathcal Z}^{(1,1)}_b(z,0)=\phi\left(-z+i{Q\over 2}\right), \quad Q=\omega_1+\omega_2.
\ee
Then the function ${\mathcal Z}^{(k,p)}_b(z,m)$ can be rewritten as
\beq
{\mathcal Z}^{(k,p)}_b(z,m)=\prod_{\gamma,\delta \in \Delta(k,p,m)} {\mathcal Z}^{(1,1)}_b\left({1\over k}(z+ib^{-1}\delta+ib\gamma);\omega_1,\omega_2\right),
\ee
where $\Delta(k,p,m)=\{\gamma,\delta\in\mathbb{Z}, 0\leq \gamma,\delta< k, \gamma-r\delta \equiv m\,
{\rm mod}\, k$\}. Now we find relation between $\gamma_M(z,m)$ for $\omega_1=b^{-1}$, $\omega_2=b$,
and ${\mathcal Z}^{(k,p)}_b(z,m)$. First, reminding the connection with $\gamma$-function \eqref{gamone},
$$
\phi(z)=\gamma\left({Q\over 2}-iz\right)^{-1},
$$
we see that
$$
{\mathcal Z}^{(1,1)}_b(z,0)=\gamma\left(Q+iz\right)^{-1}.
$$
Recalling the reflection formula (\ref{otr}), we can write
\beq
\gamma(z)=e^{\pi i B_{2,2}(z;b,b^{-1})}{\mathcal Z}^{(1,1)}_b(iz,0).
\ee
Taking the product of both sides of this relation over the lattice points $\Delta(k,p,m)$
and reminding (\ref{gampk}), we obtain
\beq
\gamma_M(z,m)=e^{\pi i A(z,m)} {\mathcal Z}^{(k,p)}_b(z,m),
\ee
where
\beq
A(z,m)=\sum_{\delta=0}^{k-1} B_{2,2}\left({1\over k}(z+\omega_1\delta+\omega_2[\! [ m+r\delta]\! ] );\omega_1,\omega_2\right),
\ee
$[\! [ x]\! ] \equiv x \mod k \in \Z_k$.
In fact,
\beq\label{ndel}
[\! [ m+r\delta]\! ] =m+r\delta-N(\delta)k,\quad {\rm if} \quad
N(\delta)k\leq m+r\delta< (N(\delta)+1)k,
\ee
where $N(\delta)=\left[{m+r\delta\over k}\right].$
Note that since $k$ and $p$ are relatively prime, when $\delta$ runs the values $0,\ldots,k-1$,
the function $[\! [ m+r\delta]\! ] $ also runs all these values, but in a different order.

Using equalities
\bea\label{d1}
\sum_{\delta=0}^{k-1} \delta=\sum_{\delta=0}^{k-1}[\! [ m+r\delta]\! ] ={k(k-1)\over 2},
\quad
\sum_{\delta=0}^{k-1} \delta^2=\sum_{\delta=0}^{k-1}[\! [ m+r\delta]\! ] ^2={k(k-1)(2k-1)\over 6},
\eea
one can show that
\beq
\sum_{\delta=0}^{k-1} B_{2,2}({1\over k}(z+\omega_1\delta+\omega_2[\! [ m+r\delta]\! ]);\omega_1,\omega_2)=
{1\over k}B_{2,2}(z;\omega_1,\omega_2)+F(m),
\ee
where
\beq\label{asd}
F(m)=-{1\over 2k}-{k\over 2}+1+{2\over k^2}\sum_{\delta=0}^{k-1}\delta [\! [ m+r\delta]\! ] .
\ee
Equalities \eqref{d1} follow from the remark given after the statement (\ref{ndel}). Thus we have
\beq\label{gamdim}
\gamma_M(z,m)=e^{{\pi i\over k}B_{2,2}(z;\omega_1,\omega_2)}e^{\pi i F(m)}{\mathcal Z}^{(k,p)}_b(iz,m).
\ee
Let us now show that the function $F(m)$ is connected with the Dedekind sum as follows
\beq\label{Fm}
e^{\pi i F(m)}=e^{{\pi i\over 2}\left(1-{1\over k}\right)}e^{2\pi i S(r,k)}e^{{\pi i p\over k}m(m+1)}(-1)^m e^{-{\pi i\over k}m}.
\ee
First, write $F(m)$ in the form
\bea \nonumber &&
F(m)={1\over 2}-{1\over 2k}+2
\sum_{\delta=0}^{k-1}{\delta\over k}
\left({m+r\delta\over k}-\left[{m+r\delta\over k}\right]-{1\over 2}\right)
\\ && \makebox[2em]{}
= {1\over 2}-{1\over 2k}+ m\left(1-{1\over k}\right)+
2\sum_{\delta=0}^{k-1}{\delta\over k}
\left({r\delta\over k}-\left[{m+r\delta\over k}\right]-{1\over 2}\right).
\label{fmtr}\eea
We saw already that equation (\ref{Fm}) is satisfied for $m=0$. To establish it for generic $m$,
we should find $m$-dependence of  the sum in the second line of (\ref{fmtr}). Namely, we should show that
\beq
e^{2\pi i \sum_{\delta=0}^{k-1}{\delta\over k}
\left({r\delta\over k}-\left[{m+r\delta\over k}\right]-{1\over 2}\right)}=
e^{2\pi i S(r,k)}e^{{\pi i p\over k}m(m+1)}.
\ee

To see validity of this equality, we analyze for which $\delta$  the integer $\left[{m+r\delta\over k}\right]$ is different
from $\left[{r\delta\over k}\right]$. Assume that $r\delta=kN-n$, for some positive integers $N$ and $n=0,1,\ldots, k-1$.
It is clear that if $m<n$, then
$\left[{m+r\delta\over k}\right]=\left[{r\delta\over k}\right]$, and if $n\leq m$, then
$\left[{m+r\delta\over k}\right]-\left[{r\delta\over k}\right]=1$. Therefore we should find $\delta$
satisfying the relation $r\delta=kN-n$ with $n\leq m$ and sum over them.
From the condition $pr=1-ks$, we obtain $- rnp=ksn-n$, i.e. the required $\delta$'s
satisfy the condition $\delta\equiv -np \mod k$ with $n\leq m$.
Since we want to compute only the phase, this condition is sufficient for our purpose.
Collecting all terms, we see that the additional phase factor created by $m\neq 0$ is
$e^{\pi i pm(m+1)/k}$.

\end{document}